\documentclass[aps,amsmath,amssymb,floatfix,twocolumn,showpacs,amsfonts,longbibliography,pra]{revtex4-1}
\usepackage{times}
\usepackage{multirow}
\usepackage[varg]{txfonts}
\usepackage{textcomp}
\usepackage{graphicx}
\usepackage{subfigure}
\usepackage{tabu}
\usepackage{color}
\usepackage[colorlinks=true,citecolor=blue,urlcolor=blue,linkcolor=blue,hyperindex]{hyperref}
\usepackage{braket}
\usepackage{float}
\usepackage{overpic}
\usepackage{amssymb}
\usepackage{amsmath}
\usepackage{gensymb}
\usepackage{mathrsfs}
\usepackage{tikz}
\usepackage{bm}

\newcommand{\ubd}[1]{\textcolor{red}{#1}}
\newcommand{\dbd}[1]{\textcolor{green}{#1}}
\allowdisplaybreaks

\begin{document}

\title{Analysis of Pseudo-Random Number Generators in QMC-SSE Method}
\author{Dong-Xu Liu}
\affiliation{Department of Physics, Chongqing University, Chongqing 401331, China}
\author{Wei Xu}
\affiliation{Department of Physics, Chongqing University, Chongqing 401331, China}
\author{Xue-Feng Zhang}
\email{Corresponding author: zhangxf@cqu.edu.cn}
\affiliation{Department of Physics, Chongqing University, Chongqing 401331, China}

\begin{abstract}
In the quantum Monte Carlo (QMC) method, the Pseudo-Random Number Generator (PRNG) plays a crucial role in determining the computation time. However, the hidden structure of the PRNG may lead to serious issues such as the breakdown of the Markov process. Here, we systematically analyze the performance of the different PRNGs on the widely used QMC method -- stochastic series expansion (SSE) algorithm. To quantitatively compare them, we introduce a quantity called QMC efficiency that can effectively reflect the efficiency of the algorithms. After testing several representative observables of the Heisenberg model in one and two dimensions, we recommend using LCG as the best choice of PRNGs. Our work can not only help improve the performance of the SSE method but also shed light on the other Markov-chain-based numerical algorithms.
\end{abstract}
\maketitle

\section{Introduction}
The Monte-Carlo method is one of the most popular algorithms in various science and technology fields \cite{MC}, and the key idea is to produce many independent samples following the specific distribution. Meanwhile, to simulate the quantum many-body system, various quantum Monte-Carlo algorithms are invented and improved \cite{QMC}, and one of the most popular methods is the stochastic series expansion (SSE) method \cite{SSE_Sandvik_0,SSE_Sandvik_1,SSE_Sandvik_2,SSE_Sandvik_3}. The target distribution is the partition function represented with a series expanded form:
\begin{equation}
    Z=\mathrm{Tr}\left\{ e^{-\beta H} \right\}=\sum_{\alpha} \sum_{n=0}^{\infty} \frac{(-\beta)^{n}}{n!} \bra{\alpha} H^n \ket{\alpha},
\end{equation}
where $H$ is the Hamiltonian simulated and $\beta=1/T$ is the inverse of the temperature. Then, the updating processes of the SSE algorithm specifically depend on the form of the Hamiltonian. However, all the random sampling processes require the PRNG. Therefore, it is straightforwardly believed that the performance of the QMC highly relies on the goodness of PRNG. 

The PRNG can generate a lot of numbers that behave as randomly distributed \cite{RNG_1990}, and ``Pseudo" means it does not originate from the true random physical processes, such as the quantum effect, thermal fluctuation, or chaos. However, because the sequence of the random numbers $\{s_0,s_1,s_2,...,s_i,...\}$ presents very low autocorrelation, it can still be used to simulate the random processes. Usually, the PRNGs working on the computer can be separated into the following steps: (i) defining the iteration function which can calculate $s_{i+1}$ based on the constant parameters and previous number $s_i$; (ii) initializing the parameters of the iteration function and also the first number $s_0$ (sometimes equal to ``seed" which specific initial condition of the iteration function); (iii) calculating the random number with help of iteration function step by step. The PRNG is pseudo, so it has some intrinsic problems, such as the periodicity $s_{p+n}=s_{n}$ with period $p$, or sometimes there exists a ``lattice" structure \cite{RNG_1983_Latt_Stru_1}. Thus, benchmarking the PRNGs in the program becomes necessary.  

In this manuscript, we test the performance of different types of PRNG on the simulation of SSE methods. A quantity named efficiency of QMC is introduced to evaluate the performance quantitatively. After checking different variables in one- and two-dimensional Heisenberg models, we provide a table explicitly demonstrating the LCG is the best. The manuscript is organized as follows: In Sec. \ref{sec2}, we briefly review the QMC-SSE algorithm and define the efficiency of QMC; In Sec.\ref{sec3}, we discuss the chosen PRNGs and the environment parameters of the computing platforms; In Sec.\ref{sec4}, the benchmark of PRNGs are presented; In Sec.\ref{sec5}, we make a conclusion.
\begin{table*}[t]
    \centering
    \begin{tabular}{|c|c|c|c|c|c|c|c|c|c|}
    \hline
       \multirow{2}{*}{PRNG} & \multicolumn{4}{c|}{One-dimension} & \multicolumn{5}{c|}{Two-dimension} \\ \cline{2-10}
        & E & $M(10^{-2})$ &$M^2$ & $S(\textbf{Q})(10^{-3})$               & E& $M(10^{-3})$ &$M^2$ & $S(\textbf{Q})(10^{-2})$ & $W^2$\\ \hline
        SFMT-607  & 344.843(4) & -0.02(12) & 5.713(3) & 2.258(1)                 & 673.670(3) &  0.02(29) & 0.4441(3) & 4.076(1)  & 6.326(4) \\  \hline
       SFMT-1279   &\ubd{344.845(4)} & -0.03(12) & 5.710(3) & 2.259(1)          & 673.674(3) & -0.05(28) & 0.4443(3) & 4.074(1)  & 6.325(3) \\ \hline
       SFMT-2281  & 344.840(4) & 0.02(12) &  5.714(3) & 2.258(1)                & 673.674(3) &  0.25(31) & 0.4445(3) & 4.073(1)  & 6.321(4) \\ \hline
       SFMT-4253  & 344.843(4) & 0.04(12) & 5.713(3) &  2.259(1)                & 673.675(3) & -0.02(29) & 0.4449(3) & 4.076(1)  & 6.322(4) \\ \hline
       SFMT-11213  & 344.840(4) & -0.02(12) & 5.713(3) &  2.259(1)              & 673.672(3) & -0.31(32) & 0.4444(3) & 4.073(1)  & 6.316(4) \\ \hline
       SFMT-19937  & 344.841(4) & 0.05(11) & 5.710(3) &  2.259(1)               & 673.675(3) &  \ubd{0.37(31)} & 0.4445(3) & 4.075(1)  & 6.325(4) \\ \hline
       SFMT-44497  & 344.842(4) & -0.03(12) & 5.711(3) &  2.258(1)              & \dbd{673.669(3)} & -0.11(28) & 0.4447(3) & 4.074(1)  & 6.320(4) \\ \hline
       SFMT-86243  & 344.843(4) & 0.01(11) & \dbd{5.707(3)} &  \ubd{2.260(1)}   & 673.676(3) & -0.18(31) & 0.4442(3) & \ubd{4.077(1)}  & \dbd{6.314(4)} \\ \hline
       SFMT-13249  & \dbd{344.839(4)} & 0.01(12) & 5.710(3) &  2.258(1)         & 673.677(3) &  0.13(32) & 0.4443(3) & 4.074(1)  & 6.321(4) \\ \hline
       SFMT-216091  & 344.843(4) & 0.06(12) & 5.714(3) &  2.259(1)              & 673.670(3) &  0.18(26) & \dbd{0.4440(3)} & 4.073(1)  & \ubd{6.329(4)} \\ \hline
       PCG  & 344.840(4) & 0.12(11) & 5.710(3) &  2.258(1)                      & 673.671(3) &  0.24(28) & 0.4441(3) & 4.072(1)  & 6.319(3) \\ \hline
       KISS  & 344.843(4) & \dbd{-0.05(12)} & 5.713(3) &  2.260(1)              & 673.670(3) & \dbd{-0.51(30)} & 0.4438(3) & 4.072(1)  & 6.322(4) \\ \hline
       LCG  & 344.842(4) & -0.04(12) & \ubd{5.714(3)} &  2.259(1)               & 673.673(3) & -0.18(27) & 0.4441(3) & 4.072(1)  & 6.315(4) \\ \hline
       WELL-512  & 344.842(4) & 0.02(12) & 5.712(3) &  2.258(1)                 & 673.673(3) &  0.02(27) & 0.4444(2) & 4.074(1)  & 6.322(4) \\ \hline
       WELL-1024  & 344.844(4) & \ubd{0.12(11)} & 5.711(3) &  2.259(1)          & 673.674(3) &  0.01(26) & 0.4449(3) & \dbd{4.072(1)}  & 6.322(4) \\ \hline
       WELL-19937  & 344.842(4) & 0.08(12) & 5.711(3) &  2.258(1)               & \ubd{673.678(3)} & -0.26(31) & \ubd{0.4450(3)} & 4.073(1)  & 6.325(4) \\ \hline
       WELL-44497  & 344.842(4) & -0.04(12) & 5.713(3) &  \dbd{2.258(1)}        & 673.670(3) & -0.26(30) & 0.4445(3) & 4.074(1)  & 6.319(4) \\ \hline
    \end{tabular}
    \caption{The numerical results of the one- and two-dimensional Heisenberg models calculated by the QMC-SSE method. The largest and smallest values are highlighted with red and green colors, respectively.}
    \label{tab1}
\end{table*}

\section{QMC-SSE Method}\label{sec2}
For the quantum spin system, the Hamiltonian is usually constructed with the diagonal and off-diagonal operators. Taking the Heisenberg model $H=J\sum_{\langle i,j\rangle}(S_i^xS_j^x+S_i^yS_j^y+S_i^zS_j^z)$ as an example, the diagonal part is the Ising term $S_i^zS_j^z$ and the off-diagonal one is the XY term $S_i^xS_j^x+S_i^yS_j^y=\frac 1 2 (S_i^+S_i^-+h.c.)$ which can exchange the spin configurations between the nearest neighbor sites. Then, the serious expansion of the partition function can be rewritten as
\begin{equation}
    Z=\sum_{\alpha} \sum_{n=0}^{\infty} \sum_{S_n}\frac{(-\beta)^{n}}{n!} \bra{\alpha} \prod_{i=1}^n H_{a_i,b_i} \ket{\alpha},
\end{equation}
where $a_i$ and $b_i$ label the type and position of the operators, and $S_n$ represents the operator sequences. The program of SSE includes three steps: initialization, thermalization, and measurement. 

In the initialization part, the parameters of the PRNG are set, especially the seed so that the results are reproducible. Meanwhile, the information of the lattice is constructed, such as the position of the bonds, coordinates of the sites, the relative position of different sites, and so on. The configuration of the operator sequence and the status of each spin are also initialized. Most importantly, the weights of the operators are calculated, also the corresponding possibility of transferring one operator to the others. It is directly related to the updating of the operator sequence. In the computer program, the initialization part is only executed once, so the consideration of its efficiency becomes unnecessary. 

The configuration of the operator sequence living in $d+1$ dimensions is updated in both thermalization and measurement parts. The updating algorithms are different \cite{QMC_singlet,QMC_SW1,QMC_SW2}, and their aims are to produce independent samples more efficiently. In the thermalization part, the system is approaching the ground state during the updating. It can be taken as thermal annealing because the number of operators increases during the updating which is equivalent to decreasing the temperature. The thermalization part should follow a single Markov chain, so it can not be parallized.

In comparison, after reaching the ground state, the operator sequence can be distributed to several computing cores so that the measurement part can be parallized and speeded up. Meanwhile, all the physical observables are calculated in each step, and their mean value $\langle O\rangle=\sum_{\alpha}\bra{\alpha}O \exp{(-\beta H)}\ket{\alpha}/Z$ can be obtained by taking Monte-Carlo average $\langle O\rangle=\sum_{i} O_i /N_{MC}$ where $O_i$ is the value of observable at $i$th step and the total number of measuring steps is $N_{MC}$. The standard error can be estimated by using bootstrap or Jack-knife method \cite{SSE_Evertz}. Here, we consider the total energy $E=\langle H\rangle$, total magnetization $M=\langle \sum_i S_i^z\rangle$ and its square $M^2=\langle (\sum_i S_i^z)^2\rangle$, structure factor $S(\textbf{Q})=\langle\sum_{l,l'} S_l^zS_{l'}^z\exp{(-i\textbf{Q}\cdot(\textbf{r}_l-\textbf{r}_{l'}))}\rangle/N^2$ and the winding numbers $W^2$ which is related to the spin stiffness (or superfluid density in bosonic language) \cite{QMC_wind}.

The ideal updating process makes the nearest Monte-Carlo steps uncorrelated. However, in the real simulation, it is extremely hard to achieve for all the observables. To quantify the correlation between Monte-Carlo steps, the autocorrelation function is introduced as follows\cite{SSE_Evertz}:
\begin{eqnarray}
C(t) &=& \langle O_{i}O_{i+t}\rangle - \langle O_{i}\rangle\langle O_{i+t}\rangle
\end{eqnarray}
where $t$ labels the Monte-Carlo steps and is sometimes called Monte-Carlo time. Then, the normalized autocorrelation function can be defined as $\Gamma(t) = {C(t)}/{C(0)}$. Typically, the autocorrelation function is an exponential decay $exp(-t/\tau)$ at large time, so that we can take $\tau$ as the autocorrelation time. It indicates the correlation between Monte-Carlo steps will decay down to $1/e$ after $\tau$ steps updates, so that $O_i$ and $O_{i+\tau}$ can be approximately taken as Markov process or uncorrelated. However, usually, there could be more than one decay mode. Therefore, here we use the integrated autocorrelation time defined as $\tau_{int} = \frac{1}{2} + \sum_{t=1}^{n} \Gamma(t)$ \cite{SSE_Evertz}. 

The standard error of the physical observables can be reduced by increasing the number of measurement steps. However, the long-range correlation between Monte-Carlo steps can weaken the accuracy of the Monte-Carlo average. Thus, we introduce the number of the effective steps $N_{eff}=N_{MC}/{2\tau_{int}}$ where the prefactor 2 is used to guarantee $N_{eff}<N_{MC}$. The computation time per step is $\bar{T}^{\textrm{C}}=T^{\textrm{C}}_t/N_{MC}$ where $T^{\textrm{C}}_t$ is total computation time. Then we name the effective steps per second as the QMC efficiency which is defined as
\begin{eqnarray}
\eta = \frac{N_{eff}}{T^{\textrm{C}}_t}= \frac{1}{2\tau_{int}\bar{T}^{\textrm{C}}}.
\end{eqnarray}
The inverse of $\eta$ is equal to the computation time per effective step, and we can clearly find shorter autocorrelation time or computation time per step can make the QMC algorithm have higher performance. Definitely, this quantity is different while considering different observables, because their autocorrelation times are determined by different modes of updating process.
\begin{table*}[t]
    \centering
    \begin{tabular}{|c|c|c|c|c|c|c|c|c|c|}
    \hline
       \multirow{2}{*}{PRNG} & \multicolumn{4}{c|}{One-dimension} & \multicolumn{5}{c|}{Two-dimension} \\ \cline{2-10}
        & E & $M$ &$M^2$ & $S(\textbf{Q})$               & E& $M$ &$M^2$ & $S(\textbf{Q})$ & $W^2$\\ \hline
 SFMT-607    & 4.276(15) & 3.121(6) & 1.600(3) & 1.656(4)                   & 2.204(7) & 0.943(2) & 0.647(1) & 0.4150(3)  & 0.730(3)  \\  \hline
 SFMT-1279   & 4.274(14) & 3.122(6) & 1.597(3) & \ubd{1.660(4)}             & 2.197(7) & 0.943(2) & 0.648(1) & 0.4148(3)  & 0.730(3)  \\ \hline
 SFMT-2281   & 4.264(14) & \ubd{3.125(6)} & \ubd{1.601(3)} & 1.657(4)       & 2.213(7) & 0.946(2) & 0.649(1) & 0.4150(3)  & 0.731(3)  \\ \hline
 SFMT-4253   & \dbd{4.262(14)} & 3.123(6) & 1.598(3) & 1.657(4)             & 2.208(7) & 0.945(2) & 0.647(1) & 0.4151(3)  & 0.731(3)  \\ \hline
 SFMT-11213  & \ubd{4.290(14)} & 3.123(7) & 1.596(3) & 1.658(4)             & 2.195(7) & 0.944(2) & 0.648(1) & 0.4153(3)  & 0.731(3)  \\ \hline
 SFMT-19937  & 4.270(14) & 3.123(6) & 1.599(3) & 1.657(4)                   & 2.206(7) & 0.945(2) & 0.646(1) & 0.4145(3)  & 0.730(3)  \\ \hline
 SFMT-44497  & 4.275(14) & \dbd{3.118(6)} & 1.599(3) & 1.658(4)             & 2.200(7) & 0.942(2) & 0.647(1) & 0.4152(3)  & 0.731(4)  \\ \hline
 SFMT-86243  & 4.270(14) & 3.121(6) & 1.597(3) & 1.657(4)                   & \ubd{2.238(9)} & 0.943(2) & 0.648(1) & 0.4150(3)  & 0.730(3)  \\ \hline
 SFMT-13249  & 4.267(14) & 3.123(6) & 1.598(3) & 1.659(4)                   & 2.217(8) & \ubd{0.947(2)} & 0.649(1) & \ubd{0.4155(3)}  & 0.730(3)  \\ \hline
 SFMT-216091 & 4.269(14) & 3.122(6) & 1.600(3) & 1.656(4)                   & 2.212(8) & 0.944(1) & \ubd{0.650(1)} & 0.4150(3)  & 0.731(3)  \\ \hline
 PCG         & 4.272(14) & 3.121(6) & 1.598(3) & 1.658(4)                   & \dbd{2.192(6)} & 0.943(1) & 0.647(1) & 0.4154(3)  & \dbd{0.598(31)} \\ \hline
 KISS        & 4.272(14) & 3.124(7) & 1.597(3) & 1.657(4)                   & 2.202(8) & \dbd{0.941(1)} & 0.647(1) & 0.4150(3)  & \dbd{0.582(32)} \\ \hline
 LCG         & 4.265(14) & 3.122(7) & \dbd{1.595(3)} & 1.657(4)             & 2.206(8) & 0.943(2) & \dbd{0.645(1)} & 0.4151(3)  & 0.731(3)  \\ \hline
 WELL-512    & 4.263(14) & 3.120(6) & 1.601(3) & 1.656(4)                   & 2.202(7) & 0.944(2) & 0.648(1) & 0.4147(3)  & 0.729(2)  \\ \hline
 WELL-1024   & 4.267(13) & 3.124(6) & 1.599(3) & 1.659(4)                   & 2.207(7) & 0.943(1) & 0.646(1) & 0.4149(3)  & 0.731(2)  \\ \hline
 WELL-19937  & 4.285(14) & \dbd{3.118(6)} & 1.598(3) & 1.657(4)             & 2.195(7) & 0.944(2) & 0.649(1) & 0.4151(3)  & \ubd{0.732(3)}  \\ \hline
 WELL-44497  & 4.272(14) & 3.123(6) & 1.600(3) & \dbd{1.655(4)}             & 2.211(7) & 0.942(2) & 0.648(1) & \dbd{0.4146(3)}  & 0.732(3)  \\ \hline
    \end{tabular}
    \caption{The autocorrelation time of different variables of the one- and two-dimensional Heisenberg model calculated by the QMC-SSE method. The largest and smallest values are highlighted with red and green colors, respectively.}
    \label{tab2}
\end{table*}

\section{PRNG and Platform}\label{sec3}
The simplest but also fastest algorithm of PRNG is LCG: Linear Congruential Generator \cite{RNG_LCG_1} which makes use of the linear congruence function as the iteration function $s_{i+1}=f(s_{i})=\textrm{Mod}(as_i+b,p)$ with all coefficients are positive integer. It is obvious that the maximum period is $p$, and all the coefficients $a$, $b$, and $p$ should be set within certain conditions so that the maximum period $p$ can be reached \cite{RNG_LCG_2}. Meanwhile, if the period is set to be the maximum of the 64-bit integer, the LCG algorithm can be strongly boosted by using the integer overflow. The integer overflow means the integer will throw the higher digits when the result is outside of the range after the operation, and it is equivalent to the function $\textrm{Mod}()$. Although it is the fastest, the limitation of the digits of the register causes a serious problem that the period of LCG can not exceed $2^{64}-1$.

In 1998, Matsumoto and Nishimura broke this constraint, and they invented a random number generator called Mersenne-Twister (MT) which is based on the matrix linear congruential method \cite{RNG_1990, RNG_MT_1}. The most popular version is the MT-19937 which has a very long period of $2^{19937}-1$. Then, M. Saito and M. Matsumoto introduced a new version of MT named SIMD-oriented Fast Mersenne Twister (SFMT) which is twice faster than conventional MT, and its period is extended to be incredibly large $2^{216091}-1$ \cite{RNG_SFMT_1}. On the other hand, another new version of the MT method named Well Equidistributed Long-period Linear Generator (WELL) has better equal distribution and longer periods, but lower CPU time consumption \cite{RNG_WELL_1}.  The Permuted Congruential Generator (PCG) was invented by Melissa E. O’Neill in 2014, and it can provide a larger period with a smaller size register \cite{RNG_PCG_1}. PCG can be taken as an improvement of the Xorshift method which uses shift and xor register method as the transform function \cite{RNG_PCG_1}. Furthermore, we also consider another famous PRNG named KISS (Keep It Simple Stupid) which also has a long period $2^{123}$ \cite{RNG_KISS_1}.

The comparison of the PRNGs has to be implemented in the same computing environment. The CPU is chosen to be the Intel Xeon gold 6420R processor (dual-channel, 2.4GHz, 24 cores, and 35.75 MB high-speed L3 Intel Smart cache), and the total memory is 256GB
(16$\times$16GB DDR4). The operating system is Linux CentOS 7.6.1810 and the program compiler is GCC 4.8.5. The optimization flags of the compiler are set as `-O3' and `-mAVX'. The SSE algorithm is the direct loop method \cite{SSE_Sandvik_3}, and is written in C++ programming language without GPU boosting. The simulated system size of one- and two-dimensional Heisenberg models are set to be 500 and 24 $\times$ 24 under periodical boundary conditions, respectively. The inverse temperature is $\beta = \frac{1}{T} = 10$ ($\beta=50$ for calculating the winding numbers). As the free parameter of SSE, the energy shift of the diagonal vertex is set to 0.3. The number of measurement steps is as large as $10^5$ and the number of thermalization steps is half of it.

Furthermore, the Xeon processor utilizes the turbo boost technique which will change the clock frequency of each processor. To rule out its influence, we use the clock() function in $<$time.h$>$ C library to count the system time and take an average of 1000 QMC runs ( 100 QMC runs for the two-dimensional case ) with different initial seeds.
\begin{figure*}[t]
	\centering
	\includegraphics[width=0.95\textwidth]{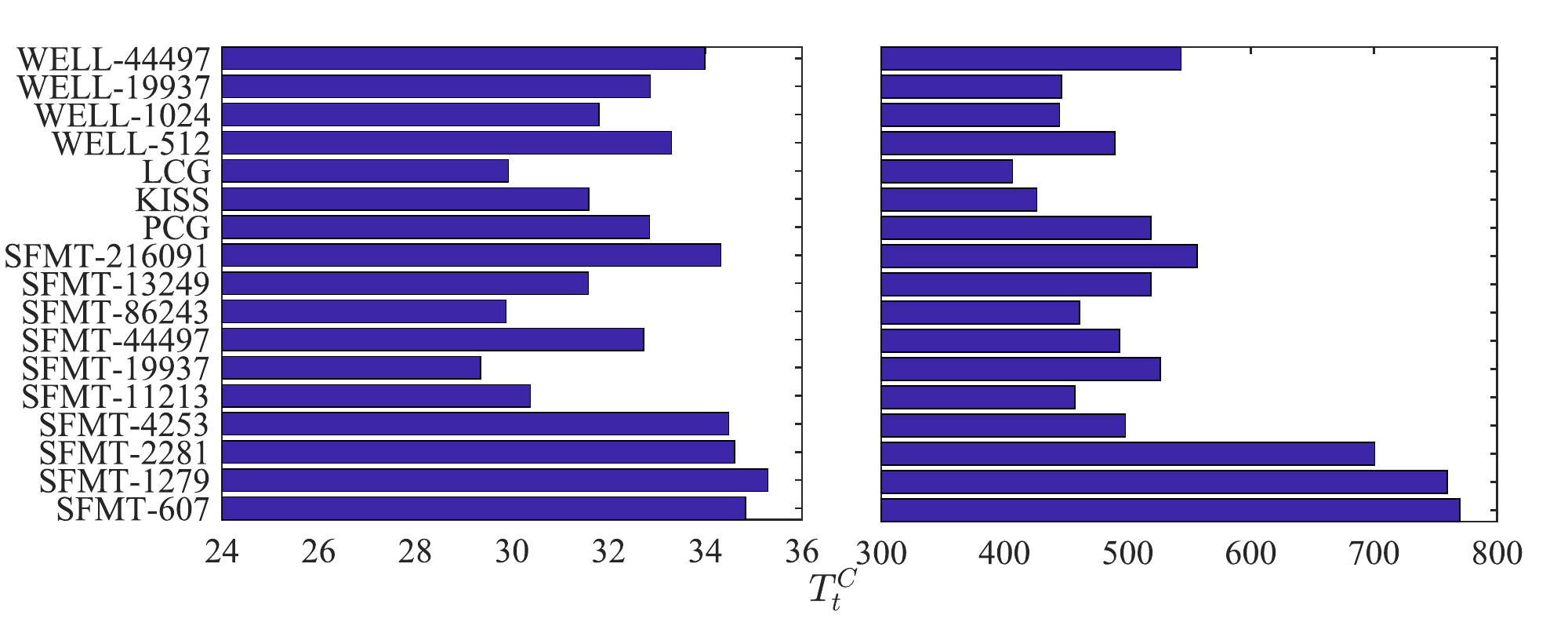}
	\caption{\label{fig1} The total computation time consumed for one- (left panel) and two-dimensional (right panel) Heisenberg model within $10^5$ Monte-Carlo steps after thermalization.}
\end{figure*}

\section{Results}\label{sec4}
Firstly, the selection of the PRNGs definitely should not cause the QMC-SSE program to produce incorrect results. As mentioned before, some PRNGs (e.g. LCG) suffer from the ``lattice" structure and they may introduce serious problems, especially in the field of cryptography. Therefore, we examine various typical observables in one- and two-dimensional Heisenberg models and list them in Table \ref{tab1}. The largest deviation values are marked with colors. Their mean values are calculated by taking an average of many independent Markov chains with different initial seeds, and the standard errors with 2$\sigma$ (95$\%$ confidence) are also shown. Here, the winding number is not considered in the one-dimensional Heisenberg model, because it is zero due to none of long-range off-diagonal order. From the numerical results, we can find there is no wrong result coming out with consideration of the statistic error. Thus, all the PRNGs listed are safe for QMC-SSE, including the LCG which is commonly used. However, it appears the SFMT algorithm with a larger period may easily cause an apparently larger deviation. 

The efficiency of the algorithm can be reflected by the autocorrelation time. As mentioned before, the updating of different variables is determined by different modes of the transfer matrix. Therefore, we list the integrated autocorrelation times of all the observables in Table \ref{tab2}. Then, we can find the autocorrelation times of the observables are strongly different and sensitive to different PRNGs, but the deviations are small which means the selection of PRNGs can not strongly affect the autocorrelation between Monte-Carlo steps. However, we notice that the PCG and KISS demonstrate much smaller autocorrelation times of winding numbers with larger standard errors. The reason can not be figured out yet, and we think it may be related to the topological property of the winding number.

The main goal of the PRNGs is to produce independent samples with low computing resource costs. It is common to take MT-19937 or LCG as PRNG when writing the QMC code. In recent decades, the PRNGs are continuously under development and their performance are strongly improved, such as the SFMT is twice faster than the MT series as mentioned on the homepage of SFMT. Different from the usual ways of thinking, the periods of PRNGs have less effect on the computation time, and it means a shorter period does not mean faster. In the QMC algorithm, the PRNGs only participate in the updating process, not involved in the measurement. Thus, the computation time is estimated by running only diagonal and loop updates after thermalization steps. In other words, the computation time affected by the PRNGs is not related to the observables. In Figure \ref{fig1}, we can find the influence of PRNGs is very serious in both dimensions. Counterintuitively, the SFMT with a smaller period costs a lot, especially in a two-dimensional system. In contrast, the SFMT-11213, 19973, and 86243 show very good performance. The LCG is widely used in SSE by A. W. Sandvik, and it is extremely good. Meanwhile, the KISS is also a considerable alternative choice. On the other hand, the computation cost of WELL series PRNGs is still fine.

Finally, we can calculate the QMC efficiency of all the observables which are listed in Table \ref{tab3}. Then, to simplify the qualification of the PRNGs, we define the benchmark score as follows: 
\begin{equation}
    \mathrm{SC}=\frac 1 n\sum_{i=1}^n \frac{\eta_{i}}{\eta_{i}^M},
\end{equation}
where $\eta_i$ is the QMC efficiency for $i$th observable with $\eta_i^M$ as the largest value among all the PRNGs, and $n$ is the number of observables considered ( e.g. $n=4$ in 1-d). Then, in one dimension, LCG, SFMT-11213, 19937 and 86243 are the best choice. In comparison, the LCG and KISS perform excellent in two dimensions. After taking an average of both dimensions, we recommend the LCG as the best PRNG in QMC-SSE. 

\section{Conclusion}\label{sec5}
The performance of the PRNGs on the QMC-SSE method is systematically analyzed in this work. The correctness of the algorithm is not ruined by the drawbacks of the PRNGs, e.g. short period, lattice structure, and so on. Meanwhile, the autocorrelation time is also less sensitive to the choice of the PRNGs, except for the winding number while using LCG and KISS. Then, we find that computation time contributes to the major impact on the performance. After introducing the QMC efficiency and benchmark score, we provide a strong recommendation on the LCG as the best PRNG in QMC-SSE method. If the lattice structure is still under worry, the KISS and SFMT-86243 would be the alternative solution. The selection of PRNGs may be highly relevant to the type of QMC, but the process of analysis we introduced here can also be borrowed as standard procedure. 

\begin{table*}[t]
    \centering
    \begin{tabular}{|c|c|c|c|c|c|c|c|c|c|c|c|c|}
    \hline
            \multirow{2}{*}{PRNG} & \multicolumn{5}{c|}{One-dimension} & \multicolumn{6}{c|}{Two-dimension} & Total\\ \cline{2-12}
            & E & $M$ &$M^2$ & $S(\textbf{Q})$ & Score  & E& $M$ &$M^2$ & $S(\textbf{Q})$ & $W^2$ & Score & Score\\ \hline
 SFMT-607    &  671  & 920   & 1794  & 1734   &   84  &   29 &   69   &   100  &   156   &   89   &   51  &   68     \\ \hline
 SFMT-1279   &  663  & 908   & 1775  & 1707   &   83  &   30 &   70   &   102  &   159   &   90   &   52  &   67     \\ \hline
 SFMT-2281   &  678  & 925   & 1805  & 1744   &   85  &   32 &   75   &   110  &   172   &   98   &   56  &   70     \\ \hline
 SFMT-4253   &  681  & 929   & 1815  & 1750   &   85  &   45 &  106   &   155  &   242   &  137   &   79  &   82     \\ \hline
 SFMT-11213  &  767  & 1054  & 2062  & 1986   &   \ubd{97}  &   50 &  116   &   169  &   263   &  150   &   86  &   \ubd{91}     \\ \hline
 SFMT-19937  &  798  & 1091  & 2130  & 2056   &  \ubd{100}  &   43 &  101   &   147  &   229   &  130   &   75  &   87     \\ \hline
 SFMT-44497  &  715  & 980   & 1911  & 1843   &   90  &   46 &  108   &   157  &   244   &  139   &   80  &   85     \\ \hline
 SFMT-86243  &  784  & 1073  & 2096  & 2020   &   \ubd{98}  &   48 &  115   &   168  &   261   &  149   &   85  &   \ubd{92}     \\ \hline
 SFMT-13249  &  742  & 1014  & 1982  & 1908   &   93  &   43 &  102   &   149  &   232   &  132   &   76  &   84     \\ \hline
 SFMT-216091 &  683  & 933   & 1822  & 1760   &   86  &   41 &   95   &   138  &   217   &  123   &   71  &   78     \\ \hline
 PCG         &  713  & 975   & 1905  & 1836   &   89  &   44 &  102   &   149  &   232   &  161   &   79  &   84     \\ \hline
 KISS        &  741  & 1014  & 1983  & 1910   &   93  &   53 &  125   &   181  &   283   &  201   &   \ubd{96}  &   \ubd{95}     \\ \hline
 LCG         &  784  & 1070  & 2095  & 2016   &   \ubd{98}  &   56 &  130   &   191  &   296   &  168   &   \ubd{97}  &   \ubd{97}     \\ \hline
 WELL-512    &  705  & 963   & 1876  & 1814   &   88  &   46 &  108   &   158  &   246   &  140   &   80  &   84     \\ \hline
 WELL-1024   &  737  & 1007  & 1966  & 1896   &   92  &   51 &  119   &   174  &   271   &  154   &   88  &   \ubd{90}     \\ \hline
 WELL-19937  &  710  & 976   & 1905  & 1837   &   89  &   51 &  119   &   172  &   270   &  153   &   88  &   89     \\ \hline
 WELL-44497  &  689  & 942   & 1838  & 1777   &   86  &   42 &   98   &   142  &   222   &  126   &   72  &   79     \\ \hline
    \end{tabular}
    \caption{The QMC efficiency and the scores of the QMC-SSE method. The recommended PRNGs are highlighted in red color.}
    \label{tab3}
\end{table*}

\section{Program Code availability}
The code used in this article has been published on GitHub at \href{https://github.com/LiuDongXu-01/QMC_RNG}{https://github.com/LiuDongXu-01/QMC$\_$RNG}. Download and decompress the code file QMC.tar.gz. Then, go into the decompressed folder. The readme file provides all the details for installing the program code and how to use it.

\section{acknowledgments}
X.-F. Z. acknowledges funding from the National Science Foundation of China under Grants  No. 12274046, No. 11874094, and No.12147102, Chongqing Natural Science Foundation under Grants No. CSTB2022NSCQ-JQX0018, Fundamental Research Funds for the Central Universities Grant No. 2021CDJZYJH-003.

\bibliographystyle{apsrev4-1}
\bibliography{ref}

\begin{thebibliography}{20}%
\makeatletter
\providecommand \@ifxundefined [1]{%
 \@ifx{#1\undefined}
}%
\providecommand \@ifnum [1]{%
 \ifnum #1\expandafter \@firstoftwo
 \else \expandafter \@secondoftwo
 \fi
}%
\providecommand \@ifx [1]{%
 \ifx #1\expandafter \@firstoftwo
 \else \expandafter \@secondoftwo
 \fi
}%
\providecommand \natexlab [1]{#1}%
\providecommand \enquote  [1]{``#1''}%
\providecommand \bibnamefont  [1]{#1}%
\providecommand \bibfnamefont [1]{#1}%
\providecommand \citenamefont [1]{#1}%
\providecommand \href@noop [0]{\@secondoftwo}%
\providecommand \href [0]{\begingroup \@sanitize@url \@href}%
\providecommand \@href[1]{\@@startlink{#1}\@@href}%
\providecommand \@@href[1]{\endgroup#1\@@endlink}%
\providecommand \@sanitize@url [0]{\catcode `\\12\catcode `\$12\catcode
  `\&12\catcode `\#12\catcode `\^12\catcode `\_12\catcode `\%12\relax}%
\providecommand \@@startlink[1]{}%
\providecommand \@@endlink[0]{}%
\providecommand \url  [0]{\begingroup\@sanitize@url \@url }%
\providecommand \@url [1]{\endgroup\@href {#1}{\urlprefix }}%
\providecommand \urlprefix  [0]{URL }%
\providecommand \Eprint [0]{\href }%
\providecommand \doibase [0]{http://dx.doi.org/}%
\providecommand \selectlanguage [0]{\@gobble}%
\providecommand \bibinfo  [0]{\@secondoftwo}%
\providecommand \bibfield  [0]{\@secondoftwo}%
\providecommand \translation [1]{[#1]}%
\providecommand \BibitemOpen [0]{}%
\providecommand \bibitemStop [0]{}%
\providecommand \bibitemNoStop [0]{.\EOS\space}%
\providecommand \EOS [0]{\spacefactor3000\relax}%
\providecommand \BibitemShut  [1]{\csname bibitem#1\endcsname}%
\let\auto@bib@innerbib\@empty
\bibitem [{\citenamefont {{Hastings}}(1970)}]{MC}%
  \BibitemOpen
  \bibfield  {author} {\bibinfo {author} {\bibfnamefont {W.~K.}\ \bibnamefont
  {{Hastings}}},\ }\href {\doibase 10.1093/biomet/57.1.97} {\bibfield
  {journal} {\bibinfo  {journal} {Biometrika}\ }\textbf {\bibinfo {volume}
  {57}},\ \bibinfo {pages} {97} (\bibinfo {year} {1970})}\BibitemShut {NoStop}%
\bibitem [{\citenamefont {Troyer}\ and\ \citenamefont {Werner}(2009)}]{QMC}%
  \BibitemOpen
  \bibfield  {author} {\bibinfo {author} {\bibfnamefont {M.}~\bibnamefont
  {Troyer}}\ and\ \bibinfo {author} {\bibfnamefont {P.}~\bibnamefont
  {Werner}},\ }\href {\doibase 10.1063/1.3225490} {\bibfield  {journal}
  {\bibinfo  {journal} {AIP Conference Proceedings}\ }\textbf {\bibinfo
  {volume} {1162}},\ \bibinfo {pages} {98} (\bibinfo {year} {2009})},\ \Eprint
  {http://arxiv.org/abs/https://pubs.aip.org/aip/acp/article-pdf/1162/1/98/11832038/98\_1\_online.pdf}
  {https://pubs.aip.org/aip/acp/article-pdf/1162/1/98/11832038/98\_1\_online.pdf}
  \BibitemShut {NoStop}%
\bibitem [{\citenamefont {Sandvik}(1992)}]{SSE_Sandvik_0}%
  \BibitemOpen
  \bibfield  {author} {\bibinfo {author} {\bibfnamefont {A.~W.}\ \bibnamefont
  {Sandvik}},\ }\href {\doibase 10.1088/0305-4470/25/13/017} {\bibfield
  {journal} {\bibinfo  {journal} {Journal of Physics A: Mathematical and
  General}\ }\textbf {\bibinfo {volume} {25}},\ \bibinfo {pages} {3667}
  (\bibinfo {year} {1992})}\BibitemShut {NoStop}%
\bibitem [{\citenamefont {Sandvik}\ \emph {et~al.}(1997)\citenamefont
  {Sandvik}, \citenamefont {Singh},\ and\ \citenamefont
  {Campbell}}]{SSE_Sandvik_1}%
  \BibitemOpen
  \bibfield  {author} {\bibinfo {author} {\bibfnamefont {A.~W.}\ \bibnamefont
  {Sandvik}}, \bibinfo {author} {\bibfnamefont {R.~R.~P.}\ \bibnamefont
  {Singh}}, \ and\ \bibinfo {author} {\bibfnamefont {D.~K.}\ \bibnamefont
  {Campbell}},\ }\href {\doibase 10.1103/PhysRevB.56.14510} {\bibfield
  {journal} {\bibinfo  {journal} {Phys. Rev. B}\ }\textbf {\bibinfo {volume}
  {56}},\ \bibinfo {pages} {14510} (\bibinfo {year} {1997})}\BibitemShut
  {NoStop}%
\bibitem [{\citenamefont {Sandvik}(1999)}]{SSE_Sandvik_2}%
  \BibitemOpen
  \bibfield  {author} {\bibinfo {author} {\bibfnamefont {A.~W.}\ \bibnamefont
  {Sandvik}},\ }\href {\doibase 10.1103/PhysRevB.59.R14157} {\bibfield
  {journal} {\bibinfo  {journal} {Phys. Rev. B}\ }\textbf {\bibinfo {volume}
  {59}},\ \bibinfo {pages} {R14157} (\bibinfo {year} {1999})}\BibitemShut
  {NoStop}%
\bibitem [{\citenamefont {Sylju\aa{}sen}\ and\ \citenamefont
  {Sandvik}(2002)}]{SSE_Sandvik_3}%
  \BibitemOpen
  \bibfield  {author} {\bibinfo {author} {\bibfnamefont {O.~F.}\ \bibnamefont
  {Sylju\aa{}sen}}\ and\ \bibinfo {author} {\bibfnamefont {A.~W.}\ \bibnamefont
  {Sandvik}},\ }\href {\doibase 10.1103/PhysRevE.66.046701} {\bibfield
  {journal} {\bibinfo  {journal} {Phys. Rev. E}\ }\textbf {\bibinfo {volume}
  {66}},\ \bibinfo {pages} {046701} (\bibinfo {year} {2002})}\BibitemShut
  {NoStop}%
\bibitem [{\citenamefont {L'Ecuyer}(1990)}]{RNG_1990}%
  \BibitemOpen
  \bibfield  {author} {\bibinfo {author} {\bibfnamefont {P.}~\bibnamefont
  {L'Ecuyer}},\ }\href {\doibase 10.1145/84537.84555} {\bibfield  {journal}
  {\bibinfo  {journal} {Communications of the ACM}\ }\textbf {\bibinfo {volume}
  {33}},\ \bibinfo {pages} {85} (\bibinfo {year} {1990})}\BibitemShut {NoStop}%
\bibitem [{\citenamefont {Ripley}\ and\ \citenamefont
  {Cox}(1983)}]{RNG_1983_Latt_Stru_1}%
  \BibitemOpen
  \bibfield  {author} {\bibinfo {author} {\bibfnamefont {B.~D.}\ \bibnamefont
  {Ripley}}\ and\ \bibinfo {author} {\bibfnamefont {D.~R.}\ \bibnamefont
  {Cox}},\ }\href {\doibase 10.1098/rspa.1983.0103} {\bibfield  {journal}
  {\bibinfo  {journal} {Proc. R. Soc. Lond. A}\ }\textbf {\bibinfo {volume}
  {389}},\ \bibinfo {pages} {197} (\bibinfo {year} {1983})}\BibitemShut
  {NoStop}%
\bibitem [{\citenamefont {{Shao}}\ \emph {et~al.}(2016)\citenamefont {{Shao}},
  \citenamefont {{Guo}},\ and\ \citenamefont {{Sandvik}}}]{QMC_singlet}%
  \BibitemOpen
  \bibfield  {author} {\bibinfo {author} {\bibfnamefont {H.}~\bibnamefont
  {{Shao}}}, \bibinfo {author} {\bibfnamefont {W.}~\bibnamefont {{Guo}}}, \
  and\ \bibinfo {author} {\bibfnamefont {A.~W.}\ \bibnamefont {{Sandvik}}},\
  }\href {\doibase 10.1126/science.aad5007} {\bibfield  {journal} {\bibinfo
  {journal} {Science}\ }\textbf {\bibinfo {volume} {352}},\ \bibinfo {pages}
  {213} (\bibinfo {year} {2016})},\ \Eprint {http://arxiv.org/abs/1603.02171}
  {arXiv:1603.02171 [cond-mat.str-el]} \BibitemShut {NoStop}%
\bibitem [{\citenamefont {Yan}\ \emph {et~al.}(2019)\citenamefont {Yan},
  \citenamefont {Wu}, \citenamefont {Liu}, \citenamefont {Sylju\aa{}sen},
  \citenamefont {Lou},\ and\ \citenamefont {Chen}}]{QMC_SW1}%
  \BibitemOpen
  \bibfield  {author} {\bibinfo {author} {\bibfnamefont {Z.}~\bibnamefont
  {Yan}}, \bibinfo {author} {\bibfnamefont {Y.}~\bibnamefont {Wu}}, \bibinfo
  {author} {\bibfnamefont {C.}~\bibnamefont {Liu}}, \bibinfo {author}
  {\bibfnamefont {O.~F.}\ \bibnamefont {Sylju\aa{}sen}}, \bibinfo {author}
  {\bibfnamefont {J.}~\bibnamefont {Lou}}, \ and\ \bibinfo {author}
  {\bibfnamefont {Y.}~\bibnamefont {Chen}},\ }\href {\doibase
  10.1103/PhysRevB.99.165135} {\bibfield  {journal} {\bibinfo  {journal} {Phys.
  Rev. B}\ }\textbf {\bibinfo {volume} {99}},\ \bibinfo {pages} {165135}
  (\bibinfo {year} {2019})}\BibitemShut {NoStop}%
\bibitem [{\citenamefont {Yan}(2022)}]{QMC_SW2}%
  \BibitemOpen
  \bibfield  {author} {\bibinfo {author} {\bibfnamefont {Z.}~\bibnamefont
  {Yan}},\ }\href {\doibase 10.1103/PhysRevB.105.184432} {\bibfield  {journal}
  {\bibinfo  {journal} {Phys. Rev. B}\ }\textbf {\bibinfo {volume} {105}},\
  \bibinfo {pages} {184432} (\bibinfo {year} {2022})}\BibitemShut {NoStop}%
\bibitem [{\citenamefont {Evertz}(2003)}]{SSE_Evertz}%
  \BibitemOpen
  \bibfield  {author} {\bibinfo {author} {\bibfnamefont {H.~G.}\ \bibnamefont
  {Evertz}},\ }\href {\doibase 10.1080/0001873021000049195} {\bibfield
  {journal} {\bibinfo  {journal} {Advances in Physics}\ }\textbf {\bibinfo
  {volume} {52}},\ \bibinfo {pages} {1} (\bibinfo {year} {2003})},\ \Eprint
  {http://arxiv.org/abs/https://doi.org/10.1080/0001873021000049195}
  {https://doi.org/10.1080/0001873021000049195} \BibitemShut {NoStop}%
\bibitem [{\citenamefont {Pollock}\ and\ \citenamefont
  {Ceperley}(1987)}]{QMC_wind}%
  \BibitemOpen
  \bibfield  {author} {\bibinfo {author} {\bibfnamefont {E.~L.}\ \bibnamefont
  {Pollock}}\ and\ \bibinfo {author} {\bibfnamefont {D.~M.}\ \bibnamefont
  {Ceperley}},\ }\href {\doibase 10.1103/PhysRevB.36.8343} {\bibfield
  {journal} {\bibinfo  {journal} {Phys. Rev. B}\ }\textbf {\bibinfo {volume}
  {36}},\ \bibinfo {pages} {8343} (\bibinfo {year} {1987})}\BibitemShut
  {NoStop}%
\bibitem [{\citenamefont {Hull}\ and\ \citenamefont
  {Dobell}(1962)}]{RNG_LCG_1}%
  \BibitemOpen
  \bibfield  {author} {\bibinfo {author} {\bibfnamefont {T.~E.}\ \bibnamefont
  {Hull}}\ and\ \bibinfo {author} {\bibfnamefont {A.~R.}\ \bibnamefont
  {Dobell}},\ }\href {\doibase 10.1137/1004061} {\bibfield  {journal} {\bibinfo
   {journal} {SIAM Review}\ }\textbf {\bibinfo {volume} {4}},\ \bibinfo {pages}
  {230} (\bibinfo {year} {1962})}\BibitemShut {NoStop}%
\bibitem [{\citenamefont {Knuth}(2002)}]{RNG_LCG_2}%
  \BibitemOpen
  \bibfield  {author} {\bibinfo {author} {\bibfnamefont {D.~E.}\ \bibnamefont
  {Knuth}},\ }\href@noop {} {\emph {\bibinfo {title} {The Art of Computer
  Programing:Seminumerical Algorithms}}},\ Vol.~\bibinfo {volume} {2}\
  (\bibinfo  {publisher} {Addison-Wesley},\ \bibinfo {year} {2002})\BibitemShut
  {NoStop}%
\bibitem [{\citenamefont {Matsumoto}\ and\ \citenamefont
  {Nishimura}(1998)}]{RNG_MT_1}%
  \BibitemOpen
  \bibfield  {author} {\bibinfo {author} {\bibfnamefont {M.}~\bibnamefont
  {Matsumoto}}\ and\ \bibinfo {author} {\bibfnamefont {T.}~\bibnamefont
  {Nishimura}},\ }\href {\doibase 10.1145/272991.272995} {\bibfield  {journal}
  {\bibinfo  {journal} {ACM Trans. Model. Comput. Simul.}\ }\textbf {\bibinfo
  {volume} {8}},\ \bibinfo {pages} {3–30} (\bibinfo {year}
  {1998})}\BibitemShut {NoStop}%
\bibitem [{\citenamefont {Saito}\ and\ \citenamefont
  {Matsumoto}(2008)}]{RNG_SFMT_1}%
  \BibitemOpen
  \bibfield  {author} {\bibinfo {author} {\bibfnamefont {M.}~\bibnamefont
  {Saito}}\ and\ \bibinfo {author} {\bibfnamefont {M.}~\bibnamefont
  {Matsumoto}},\ }\href@noop {} {\enquote {\bibinfo {title} {Simd-oriented fast
  mersenne twister: a 128-bit pseudorandom number generator},}\ } (\bibinfo
  {year} {2008})\BibitemShut {NoStop}%
\bibitem [{\citenamefont {Panneton}\ \emph {et~al.}(2006)\citenamefont
  {Panneton}, \citenamefont {L'Ecuyer},\ and\ \citenamefont
  {Matsumoto}}]{RNG_WELL_1}%
  \BibitemOpen
  \bibfield  {author} {\bibinfo {author} {\bibfnamefont {F.}~\bibnamefont
  {Panneton}}, \bibinfo {author} {\bibfnamefont {P.}~\bibnamefont {L'Ecuyer}},
  \ and\ \bibinfo {author} {\bibfnamefont {M.}~\bibnamefont {Matsumoto}},\
  }\href {\doibase 10.1145/1132973.1132974} {\bibfield  {journal} {\bibinfo
  {journal} {ACM Trans. Math. Softw.}\ }\textbf {\bibinfo {volume} {32}},\
  \bibinfo {pages} {1–16} (\bibinfo {year} {2006})}\BibitemShut {NoStop}%
\bibitem [{\citenamefont {O'Neill}(2014)}]{RNG_PCG_1}%
  \BibitemOpen
  \bibfield  {author} {\bibinfo {author} {\bibfnamefont {M.~E.}\ \bibnamefont
  {O'Neill}},\ }\href {https://www.cs.hmc.edu/tr/hmc-cs-2014-0905.pdf}
  {\enquote {\bibinfo {title} {Pcg : A family of simple fast space-efficient
  statistically good algorithms for random number generation},}\ } (\bibinfo
  {year} {2014})\BibitemShut {NoStop}%
\bibitem [{\citenamefont {Rose}(2018)}]{RNG_KISS_1}%
  \BibitemOpen
  \bibfield  {author} {\bibinfo {author} {\bibfnamefont {G.~G.}\ \bibnamefont
  {Rose}},\ }\href {\doibase 10.1007/s12095-017-0225-x} {\bibfield  {journal}
  {\bibinfo  {journal} {Cryptography Commun.}\ }\textbf {\bibinfo {volume}
  {10}},\ \bibinfo {pages} {123–137} (\bibinfo {year} {2018})}\BibitemShut
  {NoStop}%
\end{thebibliography}%
\end{document}